\documentclass[PRB,twocolumn,showpacs,amsmath,amssymb,floatfix]{revtex4-1}
\usepackage{graphicx}
\usepackage{dcolumn}
\usepackage{bm}

\begin{document}

\title{Partially disordered antiferromagnetism and multiferroic behavior in a frustrated Ising system, CoCl$_{2}$-2SC(NH$_{2}$)$_{2}$}

\author{Eundeok Mun,$^{1,*}$ Dagmar Franziska Weickert$^{2}$, Jaewook Kim,$^{1,+}$ Brian L. Scott$^{3}$, Corneliu Miclea$^{2}$, Roman Movshovich$^{2}$, Jason Wilcox$^{4}$, Jamie Manson$^{4}$, Vivien S. Zapf$^{1}$}

\affiliation{$^{1}$National High Magnetic Field Laboratory (NHMFL) Materials Physics and Applications - Condensed Matter and Magnetic Science (MPA-CMMS), Los Alamos National
Laboratory (LANL), Los Alamos, New Mexico 87545, USA}%
\affiliation{$^{2}$MPA-CMMS, LANL, Los Alamos, New Mexico 87545, USA}%
\affiliation{$^{3}$MPA-11, LANL, Los Alamos, New Mexico 87545, USA}%
\affiliation{$^{4}$Department of Chemistry and Biochemistry, Eastern Washington University, Cheney, Washington 99004, USA}%
\affiliation{$^{*}$Now at Simon Frazer University, Burnaby, British Columbia, V5A 1S6, Canada}%
\affiliation{$^{+}$Now at Rutgers Center for Emergent Materials, Piscataway, NJ 08854, USA}%

\begin{abstract}

We investigate partially disordered antiferromagnetism in CoCl$_2$-2SC(NH$_2$)$_2$, in which a-b plane hexagonal layers are staggered along the c-axis, rather than stacked. A robust 1/3 state forms in applied magnetic fields which the spins are locked, varying neither as a function of temperature nor field. By contrast, in zero field, partial antiferromagnetic order occurs, in which free spins are available to create a Curie-like magnetic susceptibility. We report measurements of the crystallographic structure, and the specific heat, magnetization, and electric polarization down to $T$ = 50 mK and up to $\mu_{0}H$ = 60 T. The Co$^{2+}$ $S = 3/2$ spins are Ising-like and form distorted hexagonal layers. The Ising energy scale is well separated from the magnetic exchange, and both energy scales are accessible to the measurements allowing us to cleanly parameterize them. In transverse fields, a quantum Ising phase transition can be observed at 2 T. Finally we find that magnetic exchange striction induces changes in the electric polarization up to 3 $\mu$C/m$^2$ and single-ion magnetic anisotropy effects induce a much a larger electric polarization change of 300 $\mu$C/m$^2$.
\end{abstract}

\pacs{Valid PACS appear here}

\maketitle

\section{Introduction}

The puzzle of frustrated magnetic spins on a triangular lattice has intrigued the scientific community for more than half a century.\cite{Wannier50} Classical Ising spins can exist in only two states, and so requiring them to accommodate a three-fold lattice with antiferromagnetic interactions is a rich source of interesting magnetic patterns on the micro- and meso-scale. The 'up up down'  or 'up up up' arrangements of Ising spins in a triangle provides the smallest magnetic unit cells. Achieving a continuously varying magnetization in the classical Ising scenario requires an infinite series larger magnetic unit cells, one for each value of the magnetization. \cite{Nakanishi83} One solution to the frustration problem is to avoid static order altogether and form a spin liquid state. \cite{Balents10,Maeno05,Anderson73} Another is to form ordering patterns on longer length scales via long-wavelength modulations or phase segregation \cite{Bak82,Selke88,Kobayashi99,Agrestini08a,Agrestini08b,Kimber11,Prsa14,Fleck10}.  Finally, partial disorder, where some spins are locked into order and others free to exhibit Curie behavior, is seen in several triangular-lattice antiferromagnets.  Mekata \cite{Mekata77}  first described the partially-disordered antiferromagnet (PDA) on stacked antiferromagnetic triangular lattices. In the PDA state, the chains of ferromagnetically-coupled spins form a magnetic state such that two chains are anti-aligned and one is disordered and free to flip with a low energy barrier. In the predicted phase diagram, the PDA state evolves into an 'up up down' state with one third of the saturation magnetization as a function of temperature or magnetic field. Several stacked triangular lattice antiferromagnets show evidence of PDA states and some indication of 1/3 plateaus including members of the ABX$_3$ and A$_3$BB'O$_6$ families. \cite{Mekata77,Hori90,Hardy04,Niitaka01a,Niitaka01b,Agrestini08a, Agrestini08b,Fleck10,Mohapatra07,Hardy06,Nishiwaki13,Lefrancois14,Jin15}

Here we present the compound CoCl$_2$-2SC(NH$_2$)$_2$, determine its crystal structure, investigate its thermodynamic properties, and construct the magnetic phase diagram  down to 50 mK and up to 60 T. This temperature and field range allows us to investigate the phase diagram up to saturation for magnetic fields perpendicular and parallel to the Ising axis, and to quantify the Ising anisotropy. For perpendicular magnetic fields, a transverse Ising scenario applies, allowing us to observe the field-induced quantum phase transition in a transverse Ising model. The structure of this material contrasts with the classic PDA model in that the hexagonal layers in the $ab$ plane are staggered, not stacked along the $c$-axis. Each spin in a given plane lies at the center of a triangle of spins in the next plane. The hexagonal lattice is distorted, e.g. the three exchange interactions $J_1$, $J_2$ and $J_3$ in the triangle correspond to different types of bonds. 

In addition to the magnetic properties, we also investigate multiferroic behavior in this material. The coupling between magnetic and electric long range order is known as the magnetoelectric multiferroic effect in which the magnetic order is modified by electric field and/or the ferroelectricity by a magnetic field \cite{Scott79, Fiebig05}.  
Most research to date in multiferroics has focused on transition-metal oxides. Coordination compounds are an alternate route to creating
magnetoelectric multiferroic behavior \cite{Jain09,Zapf10,Zapf11,Tian14} with soft and sometimes designable lattice structures. Crystallized organic molecules of thiourea \cite{Goldsmith59},
SC(NH$_{2}$)$_{2}$, and croconic acid \cite{Horiuchi10}, H$_{2}$C$_{5}$O$_{5}$, are  examples of organic ferroelectrics. The coordination compound
NiCl$_{2}$-4SC(NH$_{2}$)$_{2}$ (DTN) \cite{Zapf11} is an example of a  thiourea-containing compound which the polar crystal structure is subject to magnetostriction by magnetically ordered spins, creating magnetoelectric coupling.  
In the compound studied here, CoCl$_2$-2SC(NH$_2$)$_2$, two thiourea molecules and two Cl atoms form a tetragonal arrangement around each Co ion. We will show that magnetostrictive distortions due to either exchange interactions or single-ion anisotropy can modify the bulk electric polarization. 

\begin{figure}%
\centering
\includegraphics[width=1\linewidth]{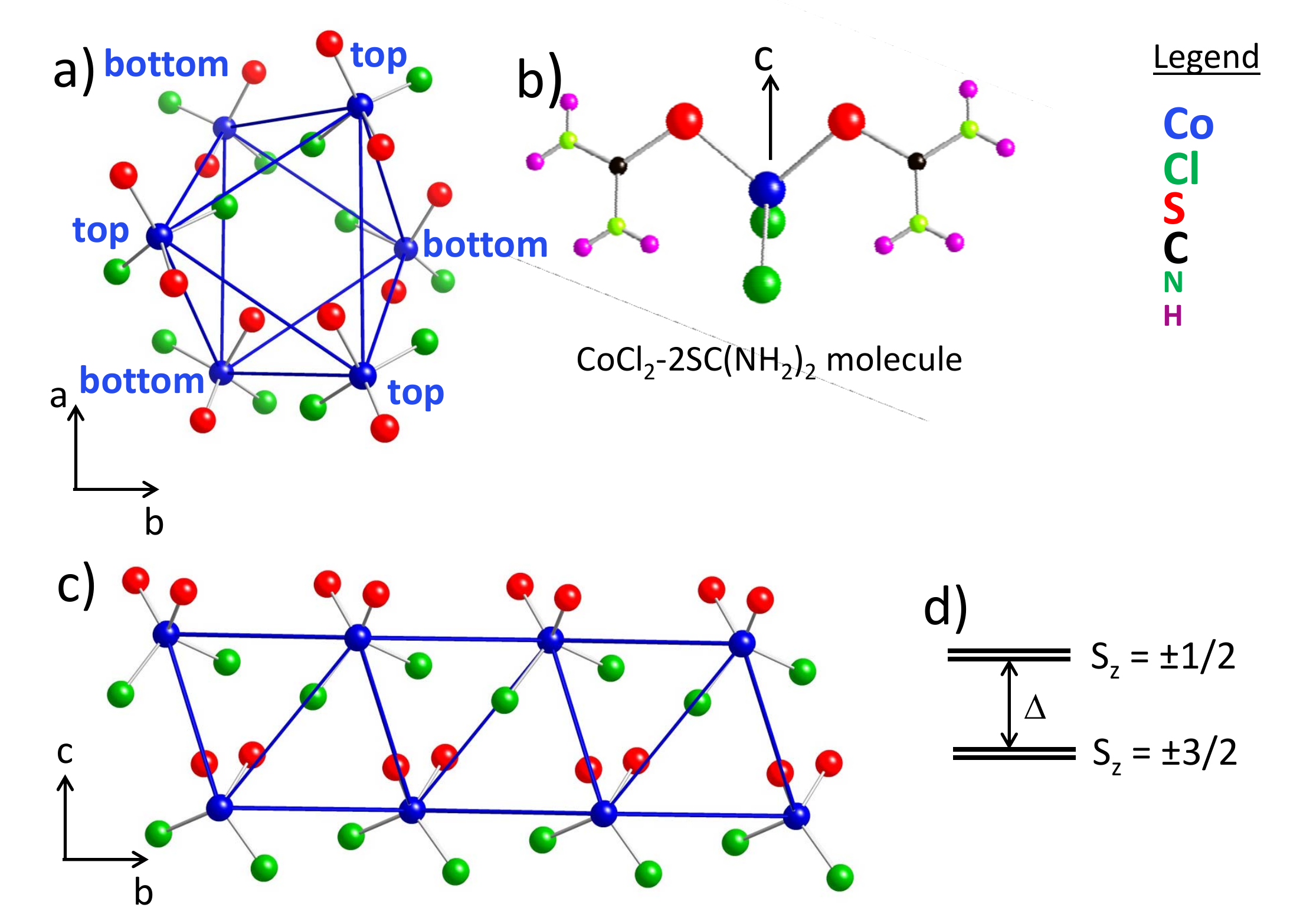}
\caption{a) Top view (along the c-axis) of the distorted hexagonal planes, showing two layers. b) A CoCl$_2$-2SC(NH$_2$)$_2$ molecule c) Side view of the stacking of the planes (viewed along the a-axis). d) Spin level diagram of the Co$^{2+}$ $\textbf{S} = 3/2$ spin levels. }
\label{crystal}%
\end{figure}%

\section{Experimental}

Single crystals of CoCl$_{2}$-2SC(NH$_{2}$)$_{2}$ were grown at EWU in ethanol solution with additional isopropanol. Initially, CoCl$_{2}$ and SC(NH$_{2}$)$_{2}$ were dissolved in warm
ethanol in separate glass beakers. After mixing the solutions, additional isopropanol was added. After slow evaporation of the solvent for two week, dark blue crystals
were obtained. Well crystallized rod-shaped crystals  were obtained with the rod-axis oriented along the crystallographic \textbf{c}-axis. 

X-ray diffraction data were collected in the MPA-11 group at LANL on a Bruker D8 diffractometer, with APEX II charge-coupled-device (CCD) detector, and an American Cryoindustries Cryocool low temperature device that cooled the sample to 140 K. The instrument was equipped with a graphite
monochromatized MoK$_{\alpha}$ X-ray source ($\lambda$ = 0.71073 \AA), and a 0.5 mm monocapillary. Crystals of CoCl$_{2}$-2SC(NH$_{2}$)$_{2}$ were mounted in a nylon cryoloop using Paratone-N oil. A hemisphere of data was collected using $\omega$ scans, with 10-second frame
exposures and 0.5$^{\texttt{o}}$ frame widths. Data collection and initial indexing and cell refinement were handled using APEX II \cite{APEX} software. Frame integration, including
Lorentz-polarization corrections, and final cell parameter calculations were carried out using SAINT+ \cite{SAINT} software. The data were corrected for absorption using redundant
reflections and the SADABS \cite{SADABS} program. Decay of reflection intensity was not observed as monitored via analysis of redundant frames. The structure was solved using Direct
methods and difference Fourier techniques. All hydrogen atom positions were idealized, and rode on the atom they were attached to. The final refinement included anisotropic
temperature factors on all non-hydrogen atoms. Structure solution, refinement, graphics, and creation of publication materials were performed using SHELXTL \cite{SHELXTL}.

Physical property measurements were performed in the MPA-CMMS group at LANL, which includes the National High Magnetic Field Laboratory Pulsed Field Facility. The temperature and magnetic field dependence of the magnetization, $M(T,H)$, were measured in a Quantum Design (QD) Physical Property Measurement System (PPMS) with a Vibrating
Sample Magnetometer (VSM) option up to $\mu_{0}H$ = 13 T and down to $T$ = 2 K. Below $T = 2$ K, $M(T,H)$ for $\textbf{H} \parallel c$ was measured with a capacitive Faraday magnetometer in an
Oxford dilution refrigerator in a 12 T superconducting magnet. In addition, $M(H)$ measurements for $\textbf{H} \parallel c$ and $\textbf{H} \perp c$ were extended up to $\mu_{0}H$ = 60 T and down to $T$ = 0.5 K in resistive pulsed magnets (10 ms rise and 40 ms decay time) driven by a capacitor bank. The pulsed-field magnetization is measured via a compensated induction coil magnetometer \cite{Detwiler00} with in-situ sample-in sample-out background subtraction.  Samples were immersed in $^3$He liquid or gas and their temperature was recorded by a resistive thermometer at zero field just prior to the pulse.  Specific heat, $C_{p}(T)$, of sample (0.78 mg) was measured by the relaxation technique down to $T$ = 50 mK in a QD PPMS with a dilution refrigerator option.

The electric polarization change, $\Delta P(H)$, was measured in pulsed magnetic fields for $\Delta\textbf{P} \perp \textbf{c}$ and for
both $\textbf{H} \parallel c$ and $\textbf{H} \perp c$  \cite{Zapf10, Zapf11}. Platinum contacts were sputtered onto the samples with a cross-sectional area of
1.2 $\times$ 1.25 mm$^{2}$ for \textbf{H} $\perp$ \textbf{c} and 1.75 $\times$ 2.6 mm$^{2}$ for \textbf{H} $\parallel$ \textbf{c}. The induced magnetoelectric currents (analogous to pyroelectric currents) due to changes in surface charge as $P$ changes with $H$ were recorded with a Stanford Research 570 current to voltage amplifier. Since the magnetoelectric current is proportional to d$\Delta P(t)$/d$t$, the measured signal is integrated as a function of time to obtain $\Delta P(H)$ with a high sensitivity due to the speed of the pulsed magnetic fields. \cite{Zapf10,Zapf11}

\section{Results}

\begin{figure}
\centering
\includegraphics[width=1\linewidth]{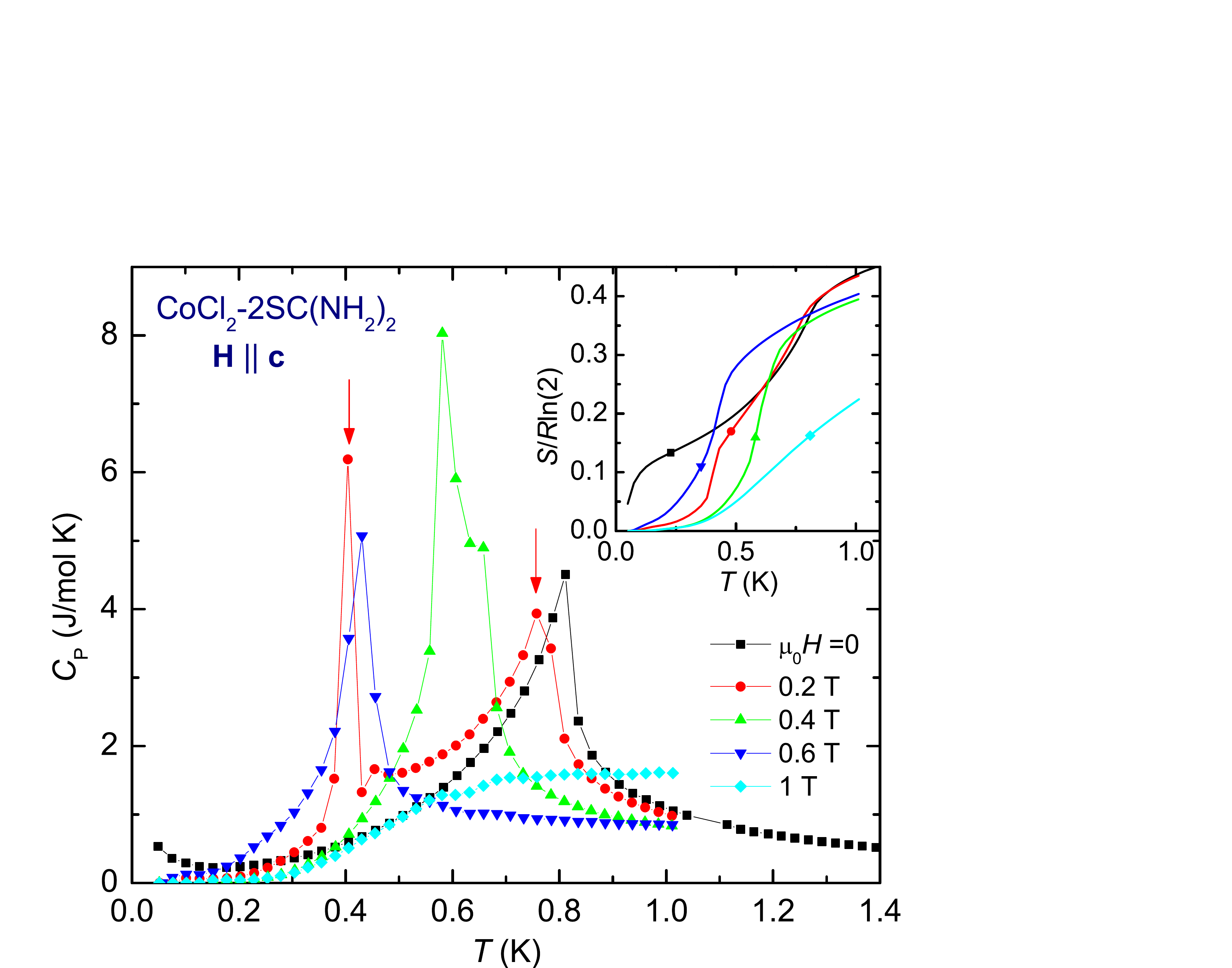}
\caption{Specific heat, $C_{p}$, data of CoCl$_{2}$-2SC(NH$_{2}$)$_{2}$, taken at various magnetic fields applied along \textbf{H} $\parallel$ \textbf{c}. Vertical arrows indicate two
phase transition temperatures for $\mu_{0}H$ = 0.2 T. Inset shows the calculated total entropy, $S(T)$, in units of $R$ln(2) by integrating $C/T$ vs $T$ at various magnetic fields.}
\label{Cp}%
\end{figure}%

\subsection{Crystal Structure}

The results of single-crystal X-ray scattering and refinement at 140 K are listed in Table \ref{table1}. We find the monoclinic structure (Cc No. 9, a = 8.199(1)\AA, b = 11.542(2)\AA, c = 10.804(2)\AA, $\beta$ = 103.587$^{\circ}$) shown in Fig.
\ref{crystal}. The Co ions form a distorted triangular lattice in the $ab$ plane that breaks spatial inversion symmetry, and allows for a net electric
polarization. The environment immediately surrounding each Co ion consists of an approximate tetrahedron with two Cl atoms and two S atoms as shown in Fig. \ref{crystal} (b). The bond
lengths involving the Co atoms are Co-Cl$_{2}$ = 2.264(7)\AA, Co-Cl$_{1}$ = 2.284(1)\AA, Co-S$_{2}$ = 2.309(9)\AA, and Co-S$_{1}$ = 2.324(0)\AA~ and the angles around the Co atoms are
Cl$_{2}$-Co-Cl$_{1}$ = 107.88(4)$^{\circ}$ and S$_{2}$-Co-S$_{1}$ = 96.85(4)$^{\circ}$.

\begin{figure*}
\centering
\includegraphics[width=0.5\linewidth]{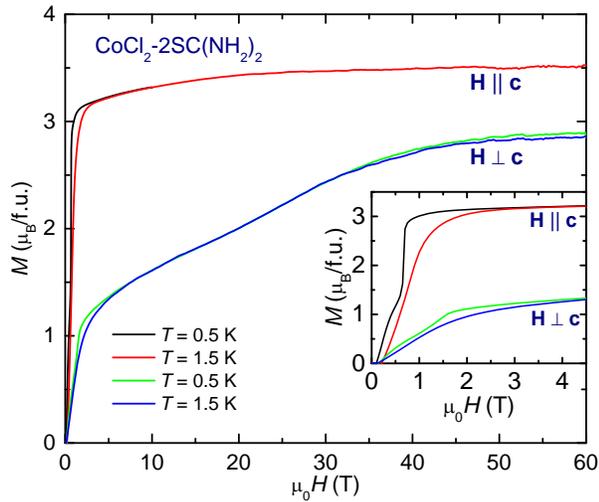}\includegraphics[width=0.5\linewidth]{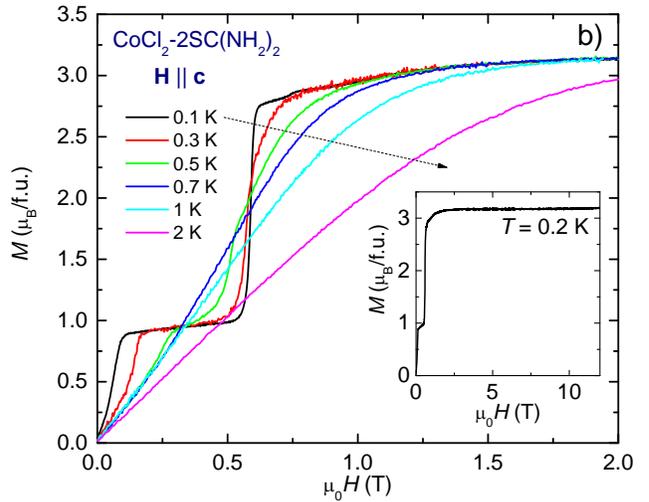}
\caption{a) High-field $M(H)$ for  \textbf{H} $\parallel$ \textbf{c} and \textbf{H} $\perp$ \textbf{c} at   $T = 0.5~$ and 1.4 K, taken in
pulsed magnetic fields up to 60 T. Vertical arrow near $H$ = 30 T indicates a slope change in d$M(H)$/d$H$. Inset shows an expanded scale at low fields. b) Magnetization isotherms, $M(H)$, of CoCl$_{2}$-2SC(NH$_{2}$)$_{2}$ for \textbf{H} $\parallel$ \textbf{c} at selected temperatures. Inset shows $M(H)$ up to 12 T taken at $T$ = 0.2 K. }
\label{MH}%
\end{figure*}%

\begin{figure}
\centering
\includegraphics[width=1\linewidth]{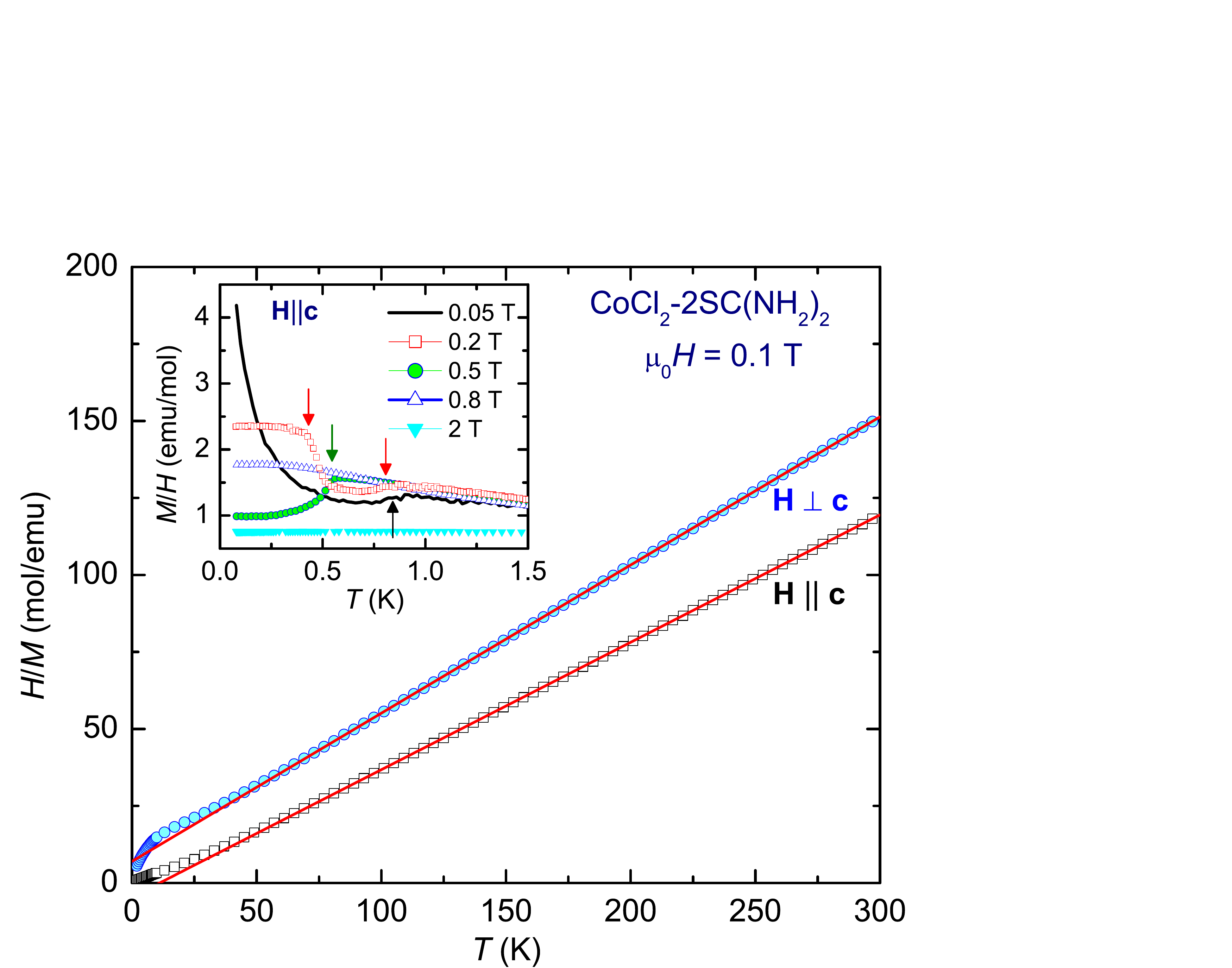}
\caption{Inverse magnetic susceptibility, $H/M(T)$, of CoCl$_{2}$-2SC(NH$_{2}$)$_{2}$ as a function of temperature for two orientations of the magnetic field; \textbf{H} $\parallel$
\textbf{c} and \textbf{H} $\perp$ \textbf{c}. Solid lines represent a Curie-Weiss fit. Inset shows magnetic susceptibility below $T$ = 1.5 K at selected magnetic fields applied along
\textbf{H} $\parallel$ \textbf{c}. Phase transition temperatures, determined by d[$(M/H)\cdot T$]/d$T$, are indicated by arrows.}
\label{MT}%
\end{figure}%

\subsection{Specific Heat}

The specific heat, $C_{p}(T,H)$ is presented in Fig. \ref{Cp}. At $H = 0$, $C(T)$ exhibits a $\lambda$-shaped anomaly at $T_N = 0.82$ K indicating a phase transition. A small upturn at the lowest temperatures, which is not seen in applied fields, may indicate a second phase transition below 50 mK. 
The anomaly centered at 0.82 K shifts to lower $T$ in applied $H$ along the  \textbf{c}-axis and vanishes by 0.6 T. For $H = 0.2$ and 0.4 T, a second anomaly is also observed as indicated by arrows in Fig. \ref{Cp}. The entropy change occurring in the phase transitions was estimated by integrating $C_{p}/T$ as a function of $T$. The total entropy change $\Delta S$ at several magnetic fields is plotted in the inset of Fig. \ref{Cp}. At these temperature,
the lattice contribution is negligible and the total entropy change is dominated by the magnetic contribution. At $H = 0$, the entropy
removed by the long range ordering is 2.1 J/mol K, which is equivalent to 40 \% of $R$ln(2). Thus the phase transitions correspond to only partial magnetic ordering.

\subsection{Magnetization vs magnetic field}

Figure \ref{MH}(a) shows the magnetization $M(H)$ up to $H = 60$ T in pulsed fields measured for  \textbf{H} $\parallel$ \textbf{c} and  \textbf{H} $\perp$ \textbf{c}, calibrated to data taken in a superconducting magnet. The data is reversible between up and down sweeps of the magnetic field, indicating no measurable heating or cooling effects during the  magnetic field pulse. The saturation magnetization by 60 T is  3.5 $\mu_{B}$/f.u. for \textbf{H} $\parallel$ \textbf{c} and  2.8 $\mu_{B}$/f.u. for  \textbf{H} $\perp$ \textbf{c}, which is close to the expected value for $\textbf{S} = 3/2$ with minimal orbital contribution. The magnetic field needed to reach saturation is highly anisotropic, being 0.6 T for  \textbf{H} $\parallel$ \textbf{c} and $\sim$40 T for  \textbf{H} $\perp$ \textbf{c}. 

Given an $\textbf{S} = 3/2$ ion with a Kramer's doublet ground state, there are two possible options for the ground state, regardless of the crystal electric field environment: an $|S^z = \pm 1/2 \rangle$ doublet ground state with an $|S^z = \pm 3/2 \rangle$ doublet excited state, or a $|S^z = \pm 3/2 \rangle$ doublet ground state with an $|S^z = \pm 1/2 \rangle$ doublet excited state. Spin-orbit interactions can also mix the different $S^z$ states but this is can be treated as a higher-order correction for a 3d ion like Co. In our system, the high-field magnetization data is consistent only with the $|S^z = \pm 3/2 \rangle$ ground state scenario. Thus for the hard axis \textbf{H} $\perp$ \textbf{c}, saturating the magnetization requires $\approx$ 40 T to overcome the anisotropy energy and rotate the spins to the hard axis, while for the easy axis \textbf{H} $\parallel$ \textbf{c}, the magnetization saturates by only 0.6 T once the antiferromagnetic order is destroyed. On the other hand, the $|S^z = \pm 1/2 \rangle$ ground state scenario can be ruled out because it would produce saturation fields that are more isotropic. It cannot account for our observed 60x difference between the saturation fields for \textbf{H} $\parallel$ \textbf{c} and \textbf{H} $\perp$ \textbf{c}. The largest energy scale in our system is the Ising anisotropy energy, and in the $|S^z = \pm 1/2 \rangle$ ground state scenario, this energy splitting needs to be overcome via the Zeeman effect in order to reach the $|S_z = 3/2 \rangle$ for \textbf{H} $\parallel$ \textbf{c} but also in order to reach the $|S_x = 3/2 \rangle$ state for \textbf{H} $\perp$ \textbf{c}. ($|S_x = 3/2 \rangle$ contains a component of $|S_z = 3/2 \rangle$). Therefore we conclude that the Co$^{2+}$ $S = 3/2$ ions have a $|S^z = \pm 3/2 \rangle$ ground state with strong Ising anisotropy and the easy axis along c.

Moving on to the low-field behavior of $M(H)$, this is shown in the inset to Fig. \ref{MH}(a) for pulsed fields for both  \textbf{H} $\parallel$ and $\perp$ \textbf{c}, at $T = 0.5$ and 1.4 K, and  also in Figure \ref{MH}b in a superconducting magnet for \textbf{H} $\parallel$ \textbf{c} down to $T =0.1$ K.  The data in the different magnets are consistent with each other. Above $T_{N} = 0.8$ K, $M(H)$ shows Brillouin-like behavior. Below $T_{N}$, $M(H)$ forms a plateau at $1/3M_{\rm{sat}}$ for $H || c$, and saturates for
$\mu_{0}H > 0.6$ T with a saturated magnetization of 3.2 $\mu_{B}$/f.u. The onset in magnetic field of the 1/3 step, and of the saturation correspond to peaks in the heat capacity shown previously. 

\subsection{Magnetic susceptibility}

The inverse magnetic susceptibility, $H/M(T)$, is plotted in Fig. \ref{MT}. A fit by a Curie-Weiss law to the
data above $T = 100$ K results in a Curie temperature $\theta_{p} = 11$ K and an effective moment $\mu_{\rm eff} = 4.4~\mu_{B}$ for \textbf{H} $\parallel$
\textbf{c} and $\theta_{p} = -15$ K and $\mu_{\rm eff} = 4.1~\mu_{B}$ for \textbf{H} $\perp$ \textbf{c}, respectively. The moments are in agreement with the expected Co$^{2+}$ ion value
with spin $\textbf{S} = 3/2$. The large anisotropy in the Curie-Weiss temperature, with a sign change from \textbf{H} $\parallel$
\textbf{c} to \textbf{H} $\perp$ \textbf{c}, indicates that the origin of the Curie-Weiss behavior is primarily the single-ion anisotropy while the exchange interactions play a smaller role \cite{Fernengel79}. This is consistent with the low values of $T_N = 0.8$ K and $H_c = 0.6$ T relative to $\theta_{p}$.

The low-temperature $M(T)/H$ curves at selected magnetic fields are plotted in the inset of Fig. \ref{MT}. The $M(T)/H$ curve at $H = 0.05$ T reveals a
kink at $T_{N} = 0.83$ K, below which $M(T)/H$ diverges with decreasing temperature. As the
magnetic field increases, $T_{N}$ shifts to lower temperatures and vanishes for $\mu_{0}H > 0.6$ T. A second anomaly in $M(T)/H$ curves is also seen for $0.1 <
\mu_{0}H < 0.4$ T (see the representative $\mu_{0}H = 0.2$ T curve in Fig. \ref{MT}). This anomaly corresponds to the second phase transitions previously shown in the heat capacity that is the onset of the "1/3" state with 1/3 of the saturation magnetization. Fig.~\ref{MT} shows that inside the 1/3 state, the magnetic susceptibility approaches a constant value as $T \rightarrow 0$, rather than diverging as it does at $H = 0$.

\begin{figure*}
\centering
\includegraphics[width=0.4\linewidth]{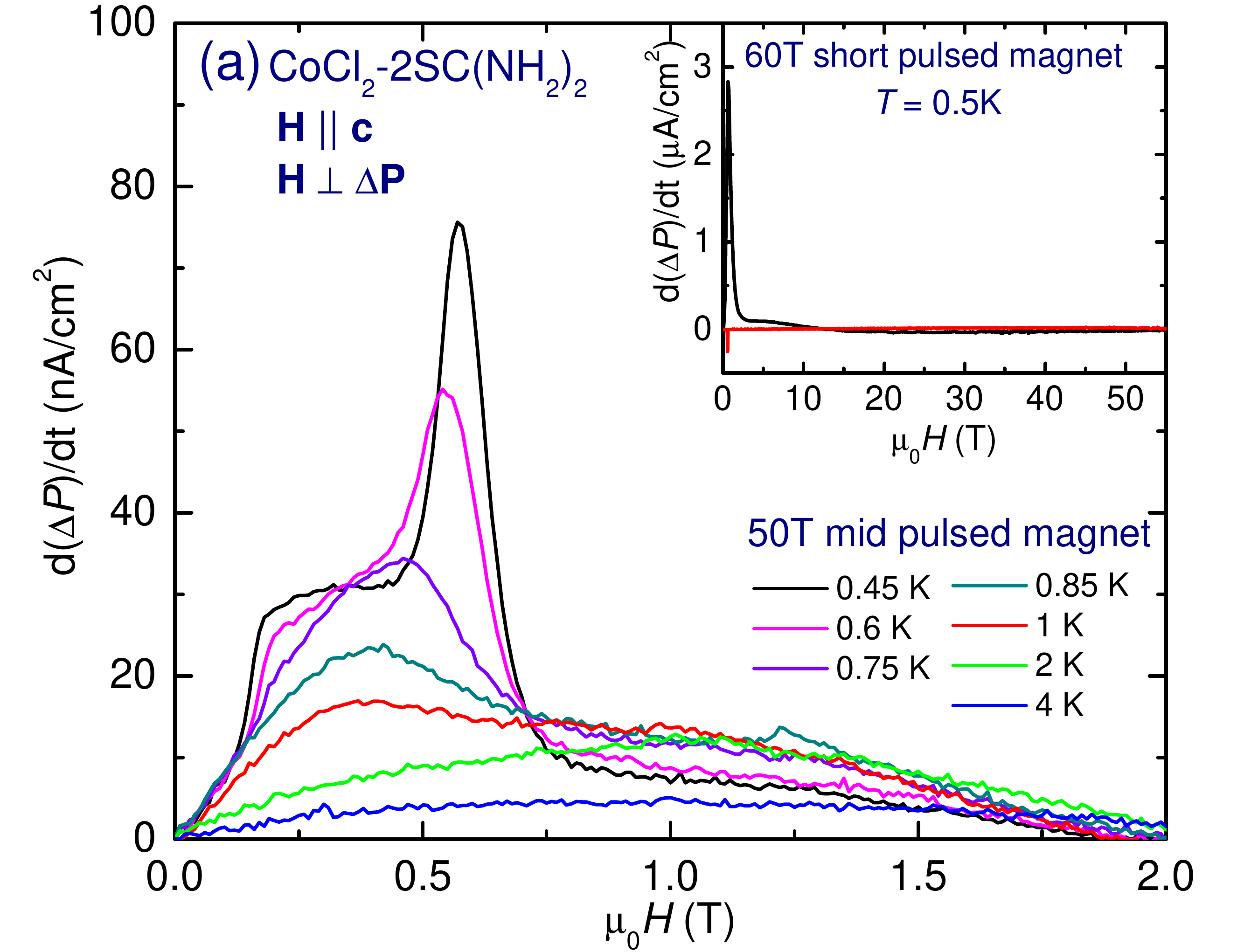}\includegraphics[width=0.4\linewidth]{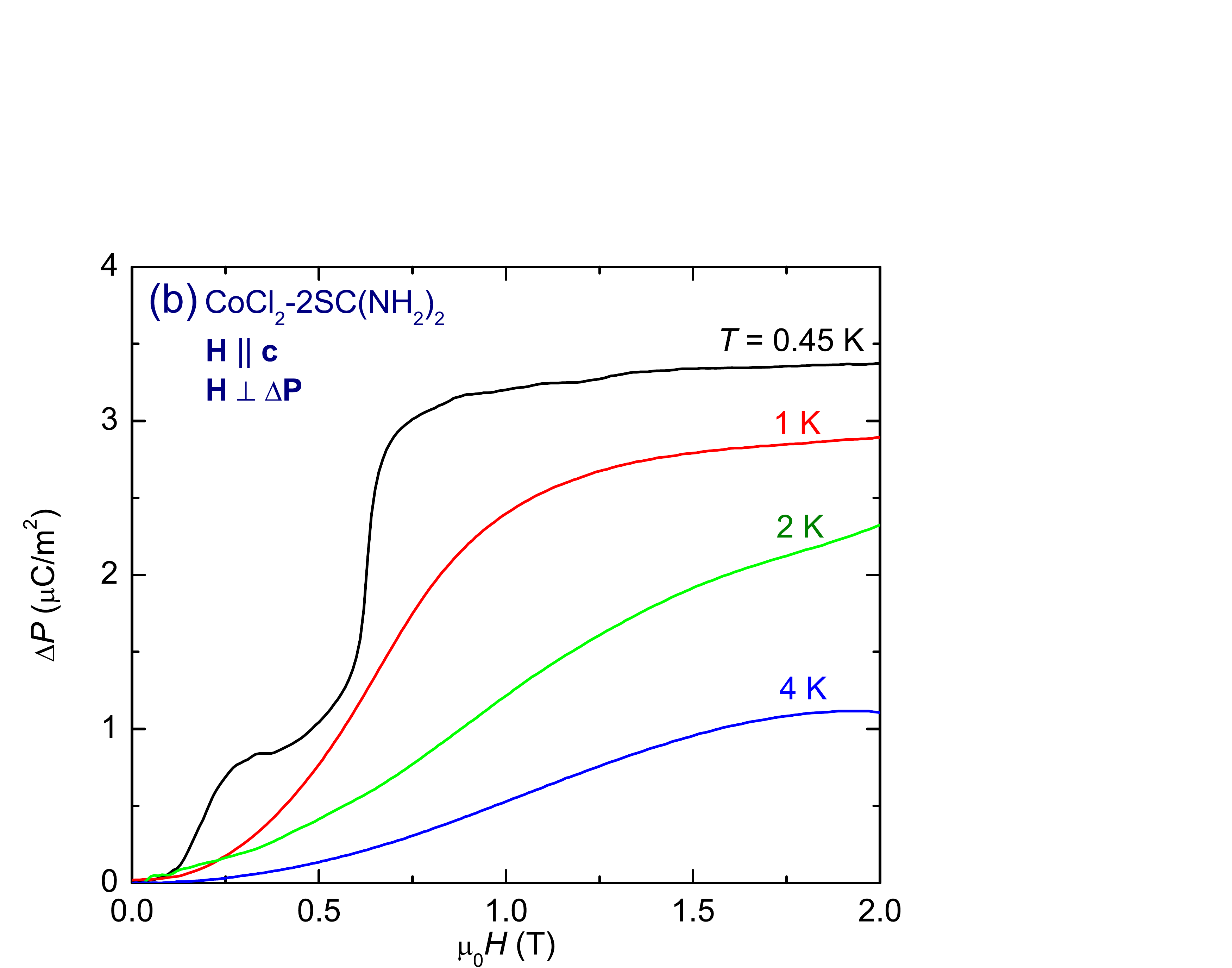}
\caption{(a) Electric polarization change with time, d$\Delta P$/d$t$, plotted as a function of magnetic field $\mu_{0}H$ at various temperatures, where \textbf{H} $\parallel$
\textbf{c} and \textbf{H} $\perp$ $\Delta$\textbf{P}. Inset shows the d$\Delta P$/d$t$ at $T = 0.48$ K up to $H = 60$ T.  (b) $\Delta P(H)$ determined
by integrating the d$\Delta P$/d$t$ as a function of time.}
\label{PH}%
\end{figure*}%

\begin{figure*}
\centering
\includegraphics[width=0.4\linewidth]{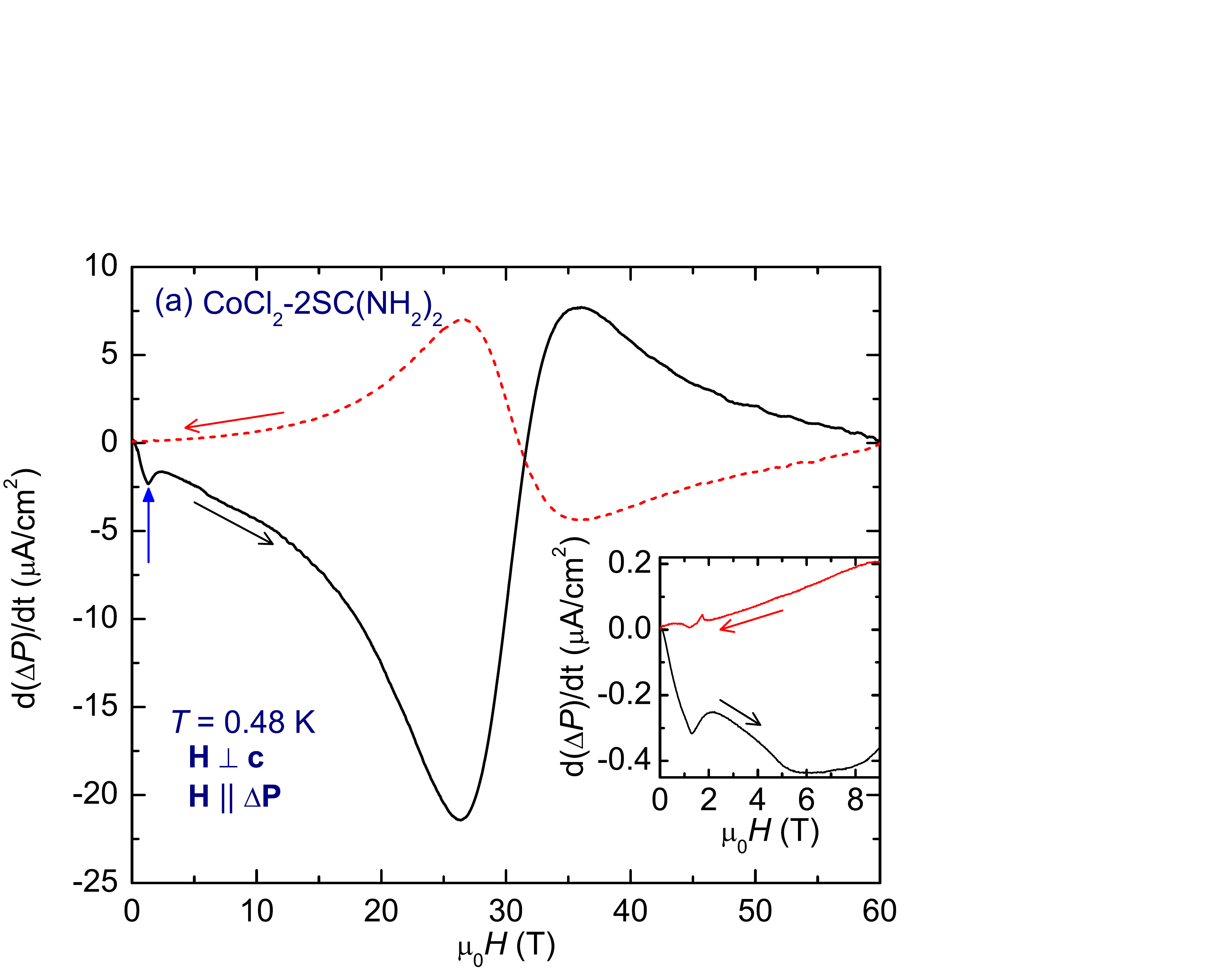}\includegraphics[width=0.4\linewidth]{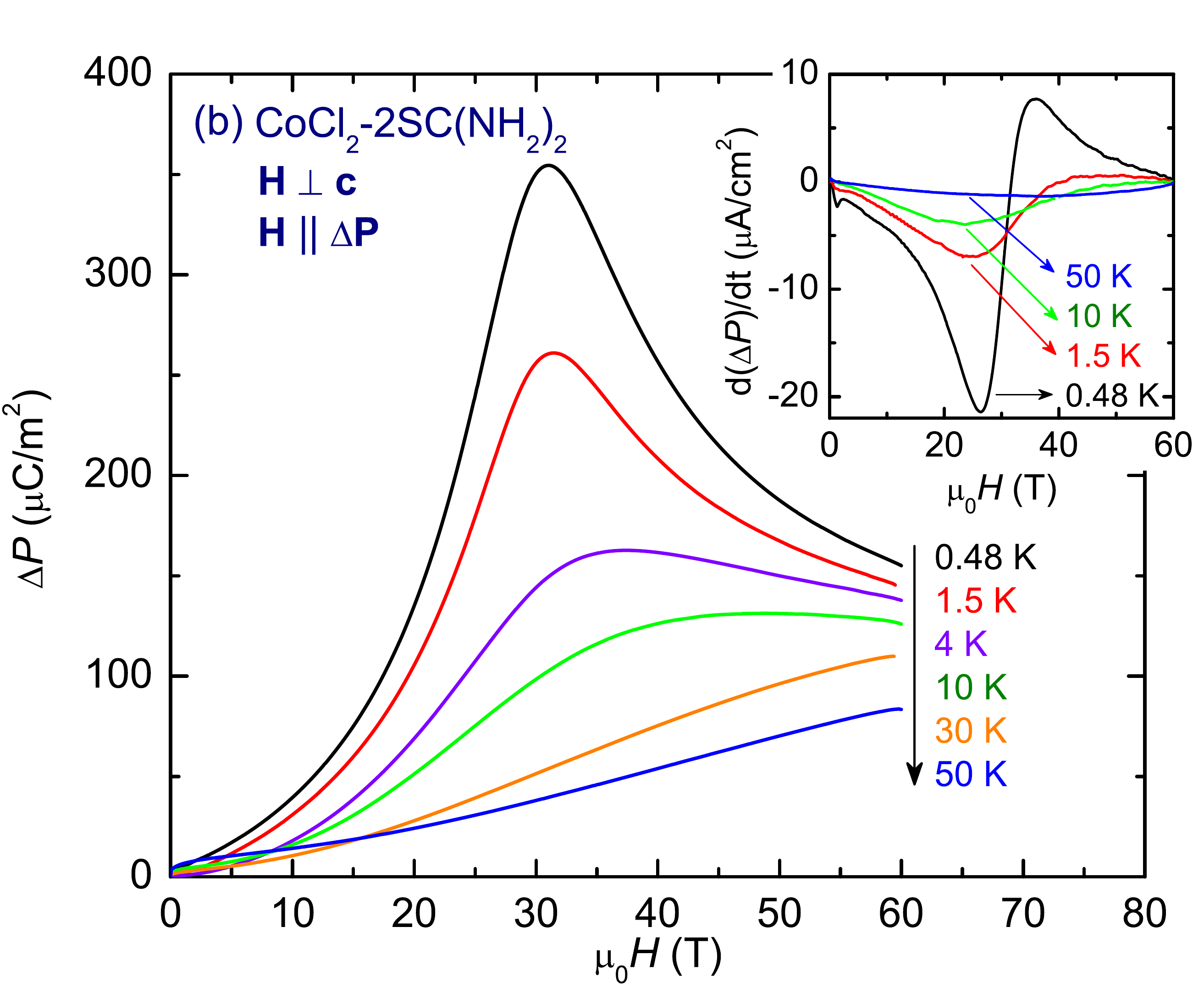}
\caption{(a) Electric polarization change, d$\Delta P$/d$t$ vs magnetic field $\mu_{0}H$, at $T$ = 0.5 K for \textbf{H} $\perp$ \textbf{c}. Inset shows an expanded plot up to
$\mu_{0}H$ = 10 T. The vertical arrow indicates a phase transition field and horizontal arrows indicate up-sweep and down-sweep. (b) $\Delta P(H)$ determined at various temperatures by
integrating the d$\Delta P$/d$t$ as a function of time. Inset shows the d$\Delta P$/d$t$ at selected temperatures.}
\label{PH1}%
\end{figure*}%

\subsection{Electric Polarization}

The electric polarization change, $\Delta P(H)$, for $\textbf{P} \perp c$ as a function of magnetic field was measured in pulsed magnetic fields up to 60 T. Consistent with the intrinsically polar crystal structure, no dependence on applied voltage up to 200 V is seen before or during measurements.  The measurement of magnetoelectric current for \textbf{H} $\parallel$ \textbf{c} and \textbf{H} $\perp$ $\Delta$\textbf{P} was performed in a slow capacitor-driven pulsed field magnet (40 ms rising and 250
ms falling time) up to 5 T, and in a faster capacitor-driven pulsed magnet (10 ms rising and 40 ms falling time) up to 60 T. The measured magnetoelectric current, d($\Delta
P$)/d$t$, induced by the polarization change of the sample and the integrated signal, $\Delta P(H)$, are plotted in Fig. \ref{PH} (a) and (b) for \textbf{H} $\parallel$ \textbf{c} and
in Fig. \ref{PH1} (a) and (b) for \textbf{H} $\perp$ \textbf{c}, respectively, where $\Delta$\textbf{P} $\perp$ \textbf{c} for both orientations of the magnetic field. The difference in magnitude of the raw d($\Delta
P$)/d$t$ data between Fig. \ref{PH} (a) and the inset is due to the different magnetic field sweep rates. This sweep-rate dependence is absent for the integrated $\Delta P(H)$ curves.

$\Delta P(H)$ in Fig. \ref{PH} shows features consistent with $M(H)$ for both magnetic field directions.   For $\textbf{H} \parallel \textbf{c}$, a 1/3 plateau is seen in $P(H)$ similar to $M(H)$, at $T = 0.45$ K. For $T\,>\,T_{N}$, $\Delta P(H)$ evolves smoothly and monotonically without any noticeable anomalies and the magnitude is immediately suppressed above $T_N$. For $\textbf{H} \perp \textbf{c}$, a kink in d($\Delta P$)/d$t$ appears at $\mu_{0}H = 1.5$ T corresponding to the suppression of long-range order (see inset) and a drastic change occurs near $\mu_{0}H = 30$ T. The integrated $P(H)$ shows a large peak at 30 T, which is just below the field where the magnetization begins to saturate. 
The same behavior is seen for both rising and falling field sweeps, except for a small hysteresis at the phase transition near $\mu_{0}H = 1.5$ T.  The observed amplitude of $\Delta P(H)$ at the phase transition is similar for both \textbf{H} $\parallel$ \textbf{c} and \textbf{H} $\perp$ \textbf{c}, whereas $\Delta P(H)$ around $\mu_{0}H$ = 30 T along \textbf{H} $\perp$ \textbf{c} is $\sim$ 100 times bigger. 

\section{Discussion}

\begin{figure}
\centering
\includegraphics[width=1\linewidth]{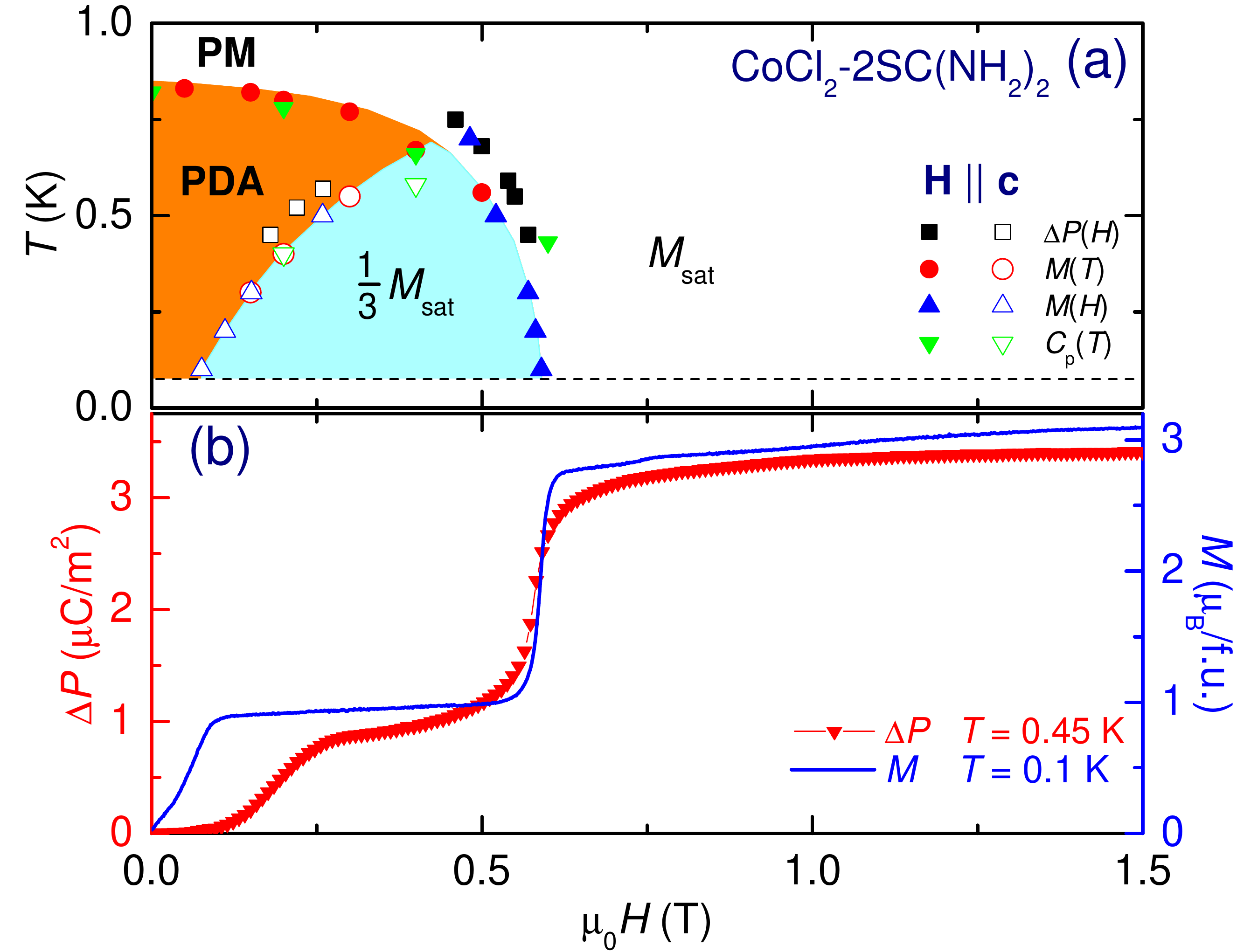}
\caption{The $H-T$ phase diagram of CoCl$_{2}$-2SC(NH$_{2}$)$_{2}$ for \textbf{H} $\parallel$ \textbf{c}, obtained from magnetization, specific heat and electric polarization
measurements. Phases are indicated as paramagnetic (PM), partially-disordered antiferromagnet (PDA), $1/3 M_{\rm sat}$, and M$_{\rm sat}$ for the phase approaching saturation.  (b) The electric polarization change, $\Delta P(H)$ left axis, at $T$ = 0.5 K and magnetization, $M(H)$ right axis, at $T$ = 0.1 K, which are the respective base temperatures of measurements.}
\label{phase}%
\end{figure}

The magnetic ordering in CoCl$_{2}$-2SC(NH$_{2}$)$_{2}$ for \textbf{H} $\parallel$ \textbf{c} is bounded by $T = 0.83$ K and $H = 0.6$ T. The phase diagram obtained by the present measurements is shown in Fig. \ref{phase} (a),
where the phase boundaries were determined from the peak position in d[$(M(T)/H)\cdot T$]/d$T$, d$M(H)$/d$H$, d$[\Delta P(H)]$/d$H$, and the peak position of $C_{p}(T)$. The phase diagram and magnetic behavior is consistent with a PDA  model. \cite{Mekata77} In the zero-field partially disordered state, the magnetization diverges at low temperatures, indicating some free spins that exhibit a Curie-behavior. In this phase, only $\sim$ 40\% of the expected entropy for fully ordering the spins is observed from the specific heat measurements. Upon applying a field, a phase transition to the 1/3 magnetization phase is observed in the heat capacity, magnetization, and electric polarization measurements. In this phase with 1/3 of the saturation magnetization, the Curie-like magnetic susceptibility disappears and the magnetization at low temperatures approaches a constant value --all spins lock into the 1/3 state. The 1/3 magnetization holds over an extended region of the $T-H$ phase diagram with no observable change in magnetization with temperature or field. This behavior is an especially pronounced example of the behavior expected from Ising spins in a PDA model. The magnetic field required to order the disordered spins and induce the 1/3 state extrapolates close to $H = 0$ at $T = 0$, indicating minimal energy barriers for flipping the free spins. Given Ising spins on a lattice with 3-fold symmetry (approximately satisfied for this compound), an 'up up down' configuration within triangles is the most likely ordering. With further increasing field the spins undergo a second field-induced phase transition out of this locked 1/3 state and then evolve towards saturation at 3 $\mu_B$. 

An important difference between this material and other PDA materials is a lack of 1-D chains in the crystal structure. The classic PDA model \cite{Mekata77} is postulated for strongly-correlated ferromagnetic chains of spins that are coordinated in a triangular motif in the perpendicular direction. In CoCl$_2$-2SC(NH$_2$)$_2$, the crystal structure does not evidence such chains - the hexagonal layers are staggered, not stacked. A spin in one plane is bonded to three spins in the next plane, forming a tetrahedron.  Thus it is a geometrically different example of partial antiferromagnetic ordering due to triangular frustration. \cite{Hori90,Hardy04,Niitaka01a,Niitaka01b,Agrestini08a, Agrestini08b,Fleck10,Mohapatra07,Hardy06,Nishiwaki13,Lefrancois14,Jin15}

We can quantify the degree of Ising-ness of the Co spins from the anisotropy between the saturation magnetic field for the easy c-axis (0.6 T) and the hard axis perpendicular to c (40 T). The scenario consistent with the data is for the Co$^{2+}$ \textbf{S} = 3/2  spins to be split into an $|S^{z} = \pm 3/2\rangle$ Kramer's protected ground state, with an excited $|S^{z} = \pm 1/2\rangle$ state separated by an energy gap $D$ (\textbf{Fig. \ref{crystal} (d)}).
For \textbf{H} $\parallel$ \textbf{c}, saturation is reached for low fields (as soon as effects of magnetic exchange interactions $J$ are overcome by 0.6 T) while saturating the magnetization for \textbf{H} $\perp$ \textbf{c} requires mixing components of $|S_{z} = \pm 3/2\rangle$ and $|S_{z} = \pm 1/2\rangle$ to achieve the $|S_{x} = 3/2\rangle$ state. Thus the gap $D$ must be closed to reach \textbf{H} $\perp$ \textbf{c} saturation near 40 T, and so we estimate $D$ $\sim$ 60 K along the c-axis, assuming $g = 2$ and $\textbf{S} = 3/2$. 

A minimal spin Hamiltonian that captures magnetic exchange and single-ion anisotropy can be written as $H = \sum_{r,v} J_{v} \textbf{S}_{\textbf{r}} \cdot \textbf{S}_{\textbf{r}+\textbf{e}_{v}} + \sum_{r}[D(S^{z}_{r})^{2} - g\mu_{B}\textbf{S}^{z}_{r} \cdot \textbf{H}]$,  where $J_{v}$ are the magnetic exchange constants, $D$ is the single-ion anisotropy due to a uniaxial crystal electric field, and $\textbf{e}_{v} = (a\hat{x}, b\hat{y}, x\hat{z})$ are the relative vectors between nearest-neighbor
Co-ions connected by magnetic exchange interactions. From the crystal structure we know that the uniaxial anisotropy axis lies along the c-axis for both Co sites in this material, and the magnetic properties show this to be the easy axis.

A comparison of the saturation fields of 0.6 T along the easy axis (needed to overcome antiferromagnetic exchange interactions) and 40 T along the hard axis (needed to overcome single-ion anisotropy) indicates that this material is in the limit of large $D/J$ with an Ising energy scale approximately 10 times larger than the magnetic exchange. A consequence of the large $D/J$ limit is that the Curie-Weiss behavior of the magnetic susceptibility is dominated by single-ion anisotropy and not by magnetic exchange. \cite{Fernengel79} Consistent with this, the Curie-Weiss temperature changes sign when the magnetic field is rotated from the easy to hard axis (-15 K vs +11 K).

For the magnetic field $\textbf{H} \perp c$, a transverse Ising scenario applies. The magnetization shows three different slopes, between $0 \leq H \leq 2$ T, $2 \leq H \leq 40$ T, and $40 \leq H \leq 60$ T.  For lower fields, we can expect the magnetization to be influenced by spin flip excitations (superposition states of $|S^z = 3/2\rangle$ and $|S^z = -3/2\rangle$). A phase transition near 2 T at 0.5 K in $M(H)$ is observed in Fig. \ref{MH}, which would be a field-induced quantum phase transition of the transverse Ising model. \cite{Ronnow05,Coldea10,Dutta12} Further investigation of transverse Ising behavior in this system is a future project requiring a detailed understanding of the exchange couplings. The final saturation near 60 T is given by the energy scale needed to overcome $D$, the gap between the ground state $|S^z = \pm 3/2\rangle$ and excited $|S^z = \pm 1/2 \rangle$ states.

Coupling between magnetism and electric polarization is shown in Fig. \ref{phase} (b). CoCl$_2$-2SC(NH$_2$)$_2$ has a polar crystal structure that allows for a net electric polarization. Thus it is a type of ferroelectric with an ordering temperature above the melting point of the crystal. Any magnetostriction-induced changes in the lattice parameters can therefore modify the electric polarization. In general, the quantities $D$ and $J$ depend on the arrangement of atoms within the crystal structure. Thus, the system will modify these through magnetostriction to minimize the combination of magnetic and crystalline energy. In the data, we observe that in the region of magnetic ordering, $\Delta P(H)$ exhibits similar magnetic field evolution as the magnetization including a 1/3 plateau and paramagnetic behavior just above $T_N$. The magnitude of $\Delta P$ is 3 $\mu$C/m$^2$ for \textbf{H} $\parallel$ \textbf{c} due to magnetic exchange striction.
Much larger values of $\Delta P(H)$ are seen for $\textbf{H} \perp c$ at higher magnetic fields up to 60 T. We can attribute these to magnetostriction driven by the single-ion $D$ term in the Hamiltonian. The value of $\Delta P(H)$  is 300 $\mu$C/m$^2$, which is within a factor of ten of the largest magnetic field-induced electric polarization changes observed in multiferroic materials \cite{Kim14,Johnson12}. The thiourea-containing compound NiCl$_2$-4SC(NH$_2$)$_2$ also shows magnetoelectric coupling due to both $D$ and $J$ terms, albeit with smaller magnitude. A detailed experimental and theoretical analysis of these effects are presented in Ref. \cite{Zapf11}.

\section{Summary}

Single crystals of a new coordination compound CoCl$_{2}$-2SC(NH$_{2}$)$_{2}$ have been synthesized. The compound has a distorted hexagonal structure and shows magnetic ordering consistent with a partially-disordered antiferromagnetic state at $H = 0$ where two spins order and one remains disordered in a triangular motif.
In applied magnetic fields, a locked 1/3 state occurs, where the magnetization remains constant at 1/3 of the saturation magnetization over a remarkably broad region in temperature and magnetic fields. The geometry of this compound differs from that of usual partially disordered antiferromagnets, since it shows staggered hexagonal planes, rather than the usual c-axis spin chains in a triangular configuration.
The Co$^{2+}$ $\textbf{S} = 3/2$ spins form a doublet ground state with $|S^z = 3/2 \rangle$ and an easy-axis anisotropy axis along the crystallographic c-axis. The well-separated energy scales for magnetic ordering and anisotropy make this a clean Ising system.
In addition, this material shows an electric polarization that is strongly coupled to magnetization with a magnetic field-induced polarization change up to 300 $\mu$C/m$^{2}$. The largest electric polarization changes in this material are produced by single-ion anisotropy effect creating magnetically-driven distortions of the Co environment.

\begin{acknowledgments}
We thank C. D. Batista and R. D. McDonald for valuable discussions. The work at Los Alamos National Lab was supported by the Laboratory Directed Research and Development program. The National High Magnetic Field Laboratory facility is funded by the US National Science Foundation through Cooperative Grant No. DMR-1157490, the State of Florida, and the US Department of Energy. The crystal design and growth at EWU was supported by the NSF under grant no. DMR-1306158.
\end{acknowledgments}

\bibliography{bibliography}
\begin{table*}
\caption{Crystallographic data}
\label{table1}%
\begin{ruledtabular}
\begin{tabular}{llll}
Formula                 &   CoCl$_{2}$-2SC(NH$_{2}$)$_{2}$     \\\hline
Formula Weight             &             282.07                             \\
$T$(K) of measurement                             & 140                               \\
Crystal system                             & monoclinic                        \\
space group                             & $Cc$                           \\
$a$ (\AA)                                     & 8.1991(15)                     \\
$b$ (\AA)                                    & 11.542(2)                         \\
$c$ (\AA)                                    & 10.804(2)                          \\
$\alpha$ ($^{\texttt{o}}$)               & 90                                \\
$\beta$ ($^{\texttt{o}}$)                   & 103.587(3)                        \\
$\gamma$ ($^{\texttt{o}}$)                    & 90                              \\
Volume (\AA$^{3}$)                           & 993.9(3)                         \\
Z                                                       & 4                                \\
Calculated $\rho$(mg/m$^{3}$)                                                         & 1.885                          \\
Crystal Size(mm$^{3}$)                                           & 0.20$\times$0.14$\times$0.14    \\
Transmission $T_{min}$, $T_{max}$                                         & 0.621, 0.710                   \\
Data Collection $\theta_{min}$, $\theta_{max}$ ($^{\texttt{o}}$)                  & 3.11, 28.44              \\
R (reflections)                                                                & 0.0309 (2274)            \\
$wR2$ (reflections)                                                           & 0.074 (2308)               \\
GOF of $F^{2}$                                                                       & 1.166                        \\
$\Delta\rho$ (max., min.) ($e$\AA$^{-3}$)                                    & 1.207, -0.891                  \\
\end{tabular}
\end{ruledtabular}
\end{table*}

\clearpage

\end{document}